\documentclass[%
 reprint,
 amsmath,amssymb,
 aps,prl
]{revtex4-2}

\usepackage[version=3]{mhchem} 
\usepackage[T1]{fontenc}       
\usepackage{dcolumn}
\usepackage{bm}
\usepackage{dcolumn}
\usepackage{amsmath,amsfonts,amssymb}
\usepackage{graphicx}
\usepackage{setspace}
\usepackage{tocloft}
\usepackage{bm}
\usepackage{lineno}
\usepackage{hyperref}
\allowdisplaybreaks

\usepackage{multirow}
\usepackage[normalem]{ulem}
\usepackage[table]{xcolor}

\begin{document}

\preprint{APS}

\title{Supercell-size scaling of moir\'e band flatness}

\author{Peilong Hong$^{1,2}$}
\email{p.l.hong@aqnu.edu.cn}
\author{Yuge Qiu$^{1}$}
\author{Wenjing Li$^{1}$}
\author{Yinying Peng$^{1}$}
\author{Yu Wang$^{1}$}
\author{Liwei Zhang$^1$}
\author{Mingfang Yi$^{1}$}
\author{Yuandi He$^1$}
\author{Peng Cheng$^1$}
\author{Wangping Cheng$^1$}
\author{Yi Liang$^3$}
\author{Guoquan Zhang$^2$}
\affiliation{$^1$School of Physics and Astronomy, Anqing Normal University, Anqing, Anhui 246133, China \\
$^2$Key Laboratory of Weak-Light Nonlinear Photonics, Ministry of Education, School of Physics, Nankai University, Tianjin 300071, China\\
$^3$School of Physical Science and Technology, Guangxi University, Nanning, Guangxi 530004, China}

\date{\today}

\begin{abstract}
In moir\'e superlattices, the band flatness governs the degree of wave localization, which is central to harnessing emergent phenomena and designing functional meta-devices. While research has focused on the \emph{magic} conditions such as magic angle and magic distance for optimal flatness, a fundamental understanding of how flatness changes with the supercell size has remained elusive. Here, we establish a universal scaling between band flatness and supercell size. Theoretically, by recognizing the statistical equivalence between structural perturbations in moir\'e superlattices and disordered systems, we introduce the Thouless number to evaluate the strength of moir\'e localization. This approach allows us to establish a scaling theory for the evolution of band flatness with the supercell size, from which an analytical expression is derived. Our full-wave simulations with one-dimensional and two-dimensional moir\'e superlattices show excellent agreement with the theoretical prediction. Our work reveals a general scaling law for moir\'e band flatness, offering a new perspective for understanding and designing moir\'e-based resonant systems.
\end{abstract}

\maketitle

The moir\'e superlattices have become an attractive platform for discovering and engineering novel quantum phases of matter~\cite{andrei2021marvels,du2023moire,oudich2024engineered}.
In condensed matter physics, moir\'e potential profoundly reconstructs the electronic band structure~\cite{andrei2021marvels}, leading to spectacular phenomena such as unconventional superconductivity~\cite{cao2018unconventional} and anomalous Hall effects~\cite{bultinck2020mechanism}. This moir\'e paradigm has rapidly extended into photonics and acoustics~\cite{du2023moire,oudich2024engineered,li2025twist}, where the moir\'e superlattices provides a highly tunable platform for engineering wave dynamics and material properties.

Central to the rich physics in moir\'e systems is the formation of flatbands~\cite{yi2022strong,wang2022intrinsic,jing2025observation,saadi2025tailoring,trushin2025flatband}. The flatbands, characterized by quenched kinetic energy, enable amplified electron-electron interaction, giving rise to strong correlation phenomena in electronic systems~\cite{andrei2021marvels}. In photonics, the flatbands with drastically reduced group velocity can enhance light-matter interaction significantly, facilitating various nonlinear optical processes such as lasing~\cite{mao2021magic,luan2023reconfigurable,ouyang2024singular} and harmonic generation~\cite{hong2022flatband,wang2024experimental}. The flatness of these bands, charactering the spectral compactness and governing the interaction strength, is thus a critical metric.

Thus far, engineering ultra-flat bands has relied on identifying finely tuned “magic” structural parameters, such as a specific twist angle (magic angle)~\cite{hu2020topological,tang2021modeling,dong2021flat} or interlayer separation (magic distance)~\cite{dong2021flat,nguyen2022magic,oudich2021photonic} in bilayer systems. At these singular conditions, a precise interplay between intralayer and interlayer coupling minimizes the bandwidth. This microscopic picture, while successful, has overshadowed a key structural degree of freedom, i.e., the size of the moir\'e supercell, defined by the number of primitive cells (N) within a single moir\'e supercell. The role of this global geometric scale in governing band flatness has remained unexplored.

Here we ask a pivotal question ``Does a universal relationship exist between the band flatness and the moir\'e supercell size, beyond the well-known dependence on magic parameters?'' Establishing such a scaling law would not only deepen the understanding of moir\'e flatband formation, but also provides a general principle for the predictive engineering of flatbands.

\begin{figure}[t]
	\centering
	\includegraphics[width=0.48\textwidth]{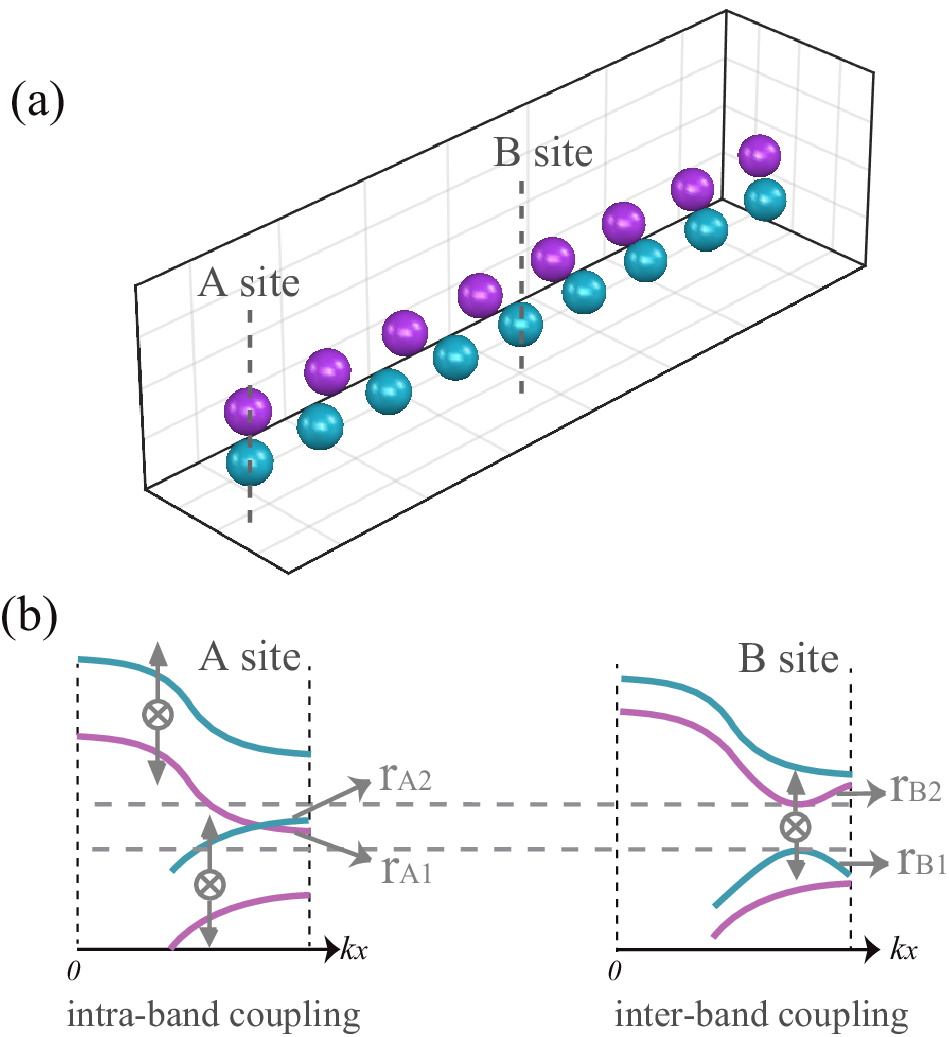}
	\caption{(a) Schematic of a one-dimensional moir\'e superlattice by overlapping two sublattices. The two sublattices exhibit different periods, leading to two extreme sites with their structures either aligned (A site) or misaligned (B site). (b) Intra-band coupling at the A site and interband coupling at the B site for an ideally infinite-size supercell. \label{Figure_1}}
\end{figure}

A moir\'e superlattice is typically formed by the overlap of two different periodic lattices, wherein the inter-lattice coupling changes continuously from one extreme site like A to the other extreme site like B in a supercell as shown in Fig.~\ref{Figure_1}(a).
Considering the ideal case that the supercell approaches infinite length, the inter-lattice coupling can be analyzed separately for the A and B sites.
As illustrated in Fig.~\ref{Figure_1}(b), the pair of sublattices aligns at the A site, leading to intra-band coupling, while a misalignment at the B site leads to interband coupling~\cite{hong2023robust}.
This unique band coupling leads a nontrivial staggered structure, i.e. the coupled bands at one site could lie within the bandgap of coupled bands at the other site, or they exhibit extremely small mode overlap with those at the other site.
Hence, the moir\'e localization could emerge at both extreme sites, leading to the formation of flatbands.
However, this theoretical picture is built on the ideal condition of infinite-size supercell. In practice, the supercell has a finite size, which means it could significantly affect the moir\'e band flatness.


To establish a theoretical framework for revealing intrinsic influence of supercell size on moir\'e band flatness, we extend the position-dependent coupled-mode theory described above. Specifically, we treat the structural transition from the A site to the B site as a perturbation introduced to an otherwise perfect A-site (or B-site) lattice. This perturbation perspective is a reminiscent of the famous Anderson localization induced by a complete disorder perturbation~\cite{lagendijk2009fifty,segev2013anderson}.
Although the moir\'e lattice and the disordered lattice are quite different, a statistical similarity between the two systems indeed exists. In detail, a disordered lattice exhibits random perturbations on an otherwise perfect lattice, indicating that the perturbed alignments are uniformly distributed across configurations ranging from the A site to the B site. Similarly, a moir\'e superlattice has interlayer alignments varying continuously from the A site to the B site, implying a uniform distribution of perturbed alignments within the A-to-B site range. This renders the structure perturbations statistically equivalent between the moir\'e lattice and the disorder lattice. Recognizing this equivalence in structural perturbations, we propose that the broadening of moir\'e bands can be characterized by the Thouless number conventionally used in the context of Anderson localization, given by~\cite{thouless1977maximum,abrahams1979scaling}
\begin{equation}
\mathrm{g}(L)=\frac{\delta E}{\Delta E}
\label{eq:g_L}
\end{equation}
Here, $\delta E$ represents the broadening of the moir\'e localized states, while $\Delta E$ denotes the mean level spacing. In this context, $\delta E$ characterizes the bandwidth of the moir\'e flatbands. To obtain the average level spacing  $\Delta E$, we consider a moir\'e lattice of supercell size $L$, which corresponds to a reciprocal space range of $2\pi/L$. According to the band-folding principle in periodic structures~\cite{joannopoulos2008molding}, the average level spacing is given by
\begin{equation}
\Delta E = \frac{\partial E}{\partial k} \cdot \frac{2\pi}{L}
\label{eq:deltaE}
\end{equation}
Note that $\frac{\partial E}{\partial k}$ is the dispersion coefficient, which can be approximated as a constant within the band of interest. For a localized state ($g\ll 1$), the Thouless number corresponds to the transmittance of a localized channel and follows an exponential scaling
\begin{equation}
g(L) = g_0 e^{-\frac{L}{\zeta}}
\label{eq:g_L2}
\end{equation}
where $g_0$ is a constant, and ${\zeta}$ represents the Thouless localization length. Combining Eqs. (1)–(3), we obtain
\begin{equation}
\delta E = \frac{\eta}{L} e^{-\frac{L}{\zeta}}
\label{eq:deltaE2}
\end{equation}
Here, $\eta = 2 \pi g_0 \frac{\partial E}{\partial k}$ is a constant. This expression captures the scaling behavior of the moir\'e band broadening with respect to the supercell size. To quantitatively characterize the flatness of the moir\'e bands, we introduce a dimensionless quantity as the flatness, i.e. the inverse normalized broadening $F = \frac{\overline{E}}{\delta E}$ with $\bar{E}$ denoting the mean energy of the moir\'e band. Hence the flatness $F$ is derived to be
\begin{equation}
F = \frac{\overline{E} }{\eta}L e^{\frac{L}{\zeta}}
\label{eq:F}
\end{equation}
Finally, the logarithmic flatness is expressed as
\begin{equation}
F_d = \ln(F) = \alpha L + \ln(L) + \beta
\label{eq:F_d}
\end{equation}
where $\alpha = \frac{1}{\zeta}$ and $\beta = \ln\left(\frac{\overline{E}}{\eta}\right)$.
Both can be viewed as constant for a flatband. Clearly, the flattening of the moir\'e band is more pronounced when its corresponding localization length ${\zeta}$ is smaller.
Given the discreteness of the moir\'e superlattice, we substitute the characteristic length $L$ with $Na$, where $a$ is the period of unperturbed lattice. Thus, the logarithmic flatness is given by 
\begin{equation}
F_{d} = \alpha_{1} N + \ln(N) + \beta_{1}
\label{eq:F_dN}
\end{equation}
where $\alpha_{1} = \alpha\cdot a$, and $\beta_{1} = \beta + \ln(a)$. Equations (6) and (7) constitute the central findings of this work. In the following, we perform full-wave simulations to verify this scaling law.


\begin{figure}[t]
	\centering
	\includegraphics[width=0.42\textwidth]{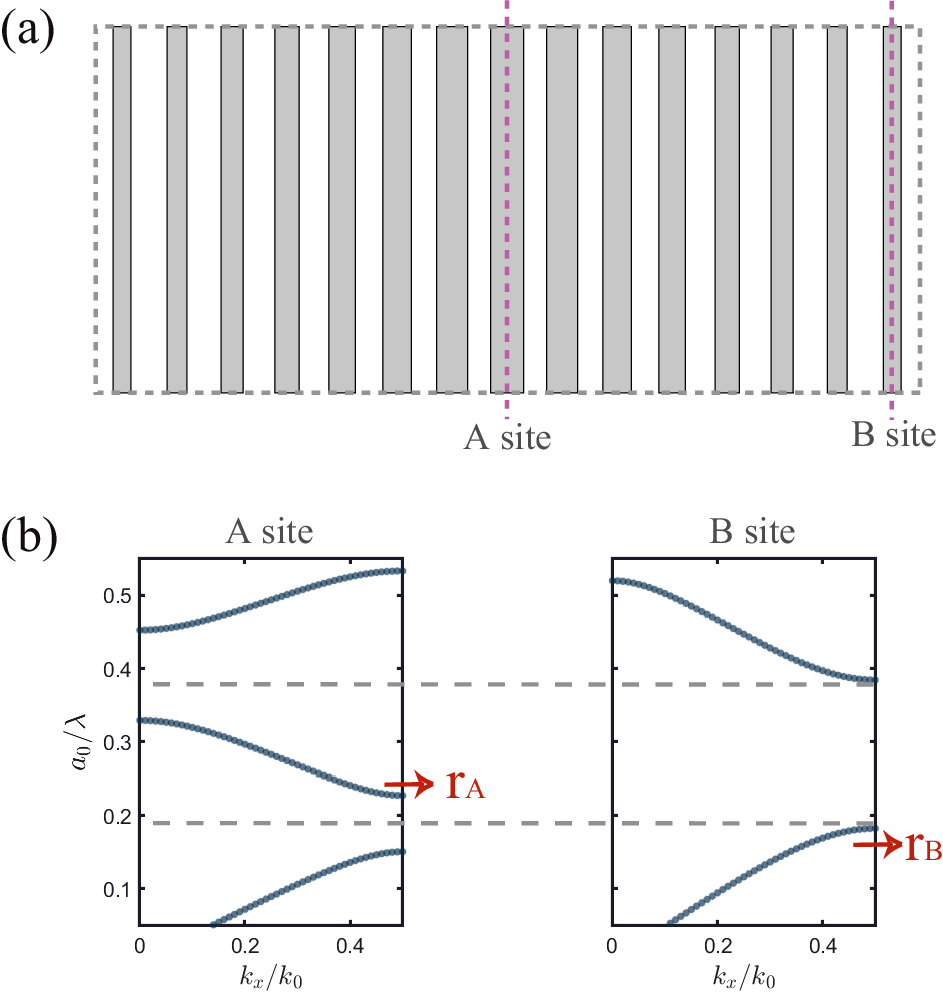}
	\caption{(a) A one dimensional moir\'e superlattice by employing continuous structure perturbations across the two extreme sites A and B.
	(b) The band structures at the A site and B site, respectively.
	The appropriate design of the structure perturbation leads to staggered band structure between the two extreme sites. \label{Figure_2}}
\end{figure}

According to the position-dependent perturbation framework, we start with a one-dimensional periodical lattice with cell length $a$, and imposes continuous structure variations within a supercell region.
We directly changes the width of the silicon bar within a cell rather than employing another lattice for perturbations, as illustrated in Fig.~\ref{Figure_2}(a).
In this case, the width $w$ of a silicon bar is reduced linearly as $w(x) = w_0-\gamma |x|$, where the coordinate $x \in [-a_m/2, a_m/2]$ with $a_m = Na$, $w_0 (= 0.6a)$ is the maximum width at A site, and $\gamma = 0.6a/a_m$ is a decay rate. 
Such a structural perturbation leads to a staggered band structure between the A site and the B site as shown in Fig.~\ref{Figure_2}(b).
Clearly, there is a band $r_A$ at the A site lies within the band gap at the B site, while a part of a band $r_B$ at the B site lies in the band gap at the A site.
Hence, wave localization could emerge at both sites, leading to the formation of moir\'e flatbands.

\begin{figure}[b]
	\centering
	\includegraphics[width=0.48\textwidth]{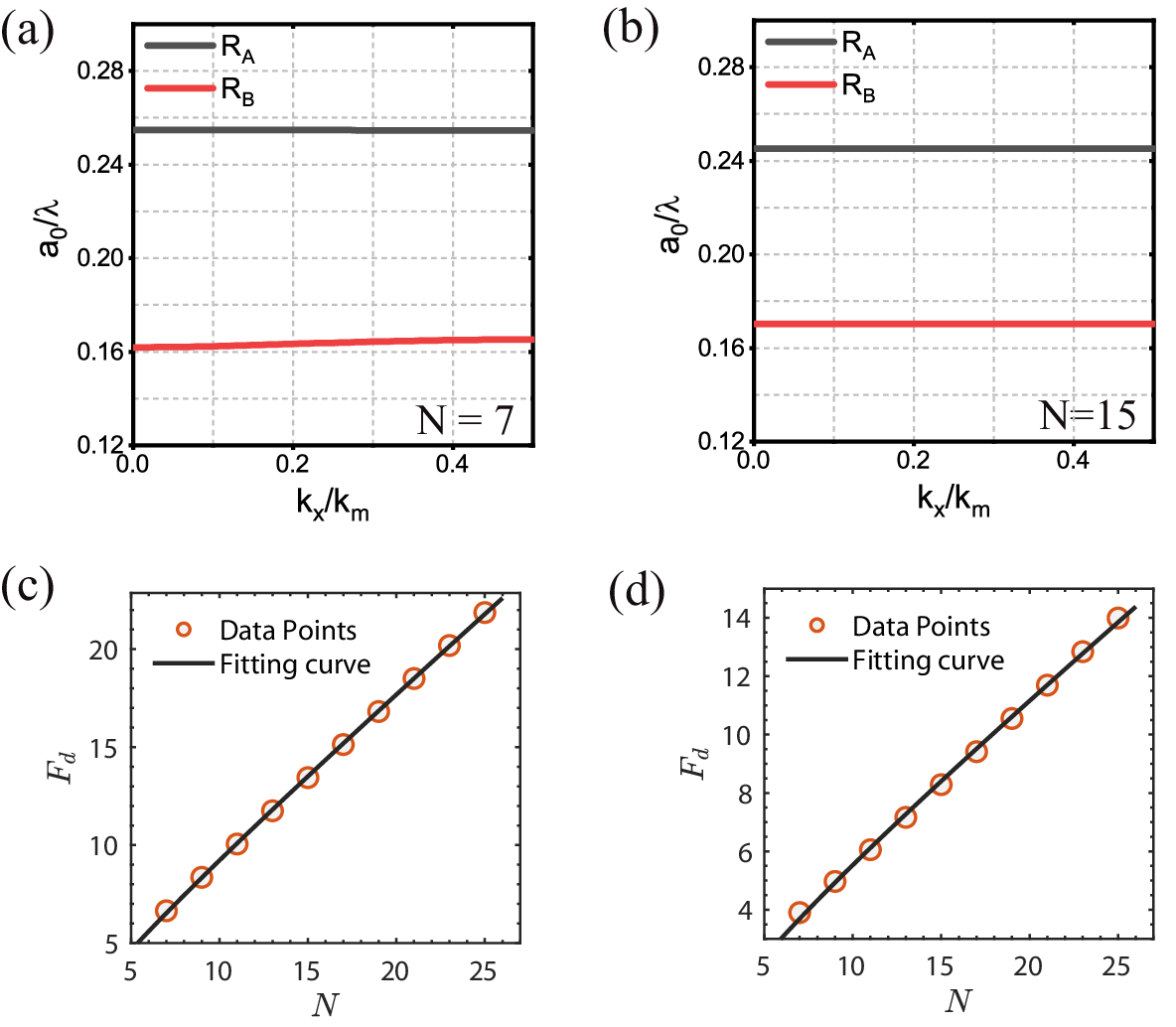}
	\caption{(a,b) Two flatbands $R_A$ and $R_B$ for different superlattice sizes ($N$ = 7 and $N$ = 15). Here, the field of $R_A$ mode is localized around $A$ site, while the field of $R_B$ mode is localized around $B$ site. It could be seen that the band flatness show a significant increase when the supercell size grows.		
(c,d) Flatness $F_d$ as a function of the supercell size $N$ for the flatband $R_{A}$ (c) and flatband $R_{B}$ (d), respectively. Here, the data points are obtained by full-wave simulation, while the curves are fitted by utilizing Eq.~\ref{eq:F_dN}.\label{Figure_3}}
\end{figure}

By utilizing finite-element method for full-wave simulation, we obtain the bands of above moir\'e lattice.
Indeed, we discover two flatbands as shown in Fig.~\ref{Figure_3}(a), which exhibit wave localization around the A and B sites, respectively.
By changing the size of the supercell from $N=7$ to $N=15$, we find that the band flatness increases, revealing a strong supercell-size effect.

To quantitatively investigate the scaling of band flatness, we calculate the normalized broadening of these moir\'e flatbands, defined as~\cite{nguyen2022magic,hong2023robust}
\begin{equation}
\frac{\delta\omega}{\bar{\omega}} = 
\frac{\max\left[\omega\left(k_{x}\right)\right] - \min\left[\omega\left(k_{x}\right)\right]}
{\dfrac{1}{k_{m}} \int_{-0.5k_{m}}^{0.5k_{m}} \omega\left(k_{x}\right) \, \mathrm{d}k_{x}}
\label{eq:relative_broadening}
\end{equation}
Accordingly, the flatness of the bands can be expressed as
\begin{equation}
F = \frac{\bar{\omega}}{\delta\omega} = 
\frac{ \dfrac{1}{k_{m}} \displaystyle\int_{-0.5k_{m}}^{0.5k_{m}} \omega(k_{x}) \, \mathrm{d}k_{x} }
{ \max[ \omega(k_{x}) ] - \min[ \omega(k_{x}) ] }
\label{eq:F_parameter}
\end{equation}
This allows us to calculate the logarithmic flatness as
\begin{equation}
F_{d} = \ln \left( \frac{ 
	\dfrac{1}{k_{m}} \int_{-0.5 k_{m}}^{0.5 k_{m}} \omega ( k_{x} ) \, \mathrm{d} k_{x}
}{
	\max [ \omega ( k_{x} ) ] - \min [ \omega ( k_{x} ) ]
} \right)
\label{eq:F_d_definition}
\end{equation}

The simulation results of the dependence of $F_d$ on $N$ are presented in Figs.~\ref{Figure_3}(c,d). For both flatbands, we fit the data points by using Eq.~\ref{eq:F_dN}, which show excellent agreement with our theoretical prediction. Notably, the fitting slop varies across different moir\'e bands, which is 0.78 for the A-site flatband and 0.49 for the B-site flatband.
Following our theory, we extract the Thouless localization length $1.28a$ for the A-site localized mode and $2.04a$ for the B-site localized mode.

\begin{figure}[t]
	\centering
	\includegraphics[width=0.48\textwidth]{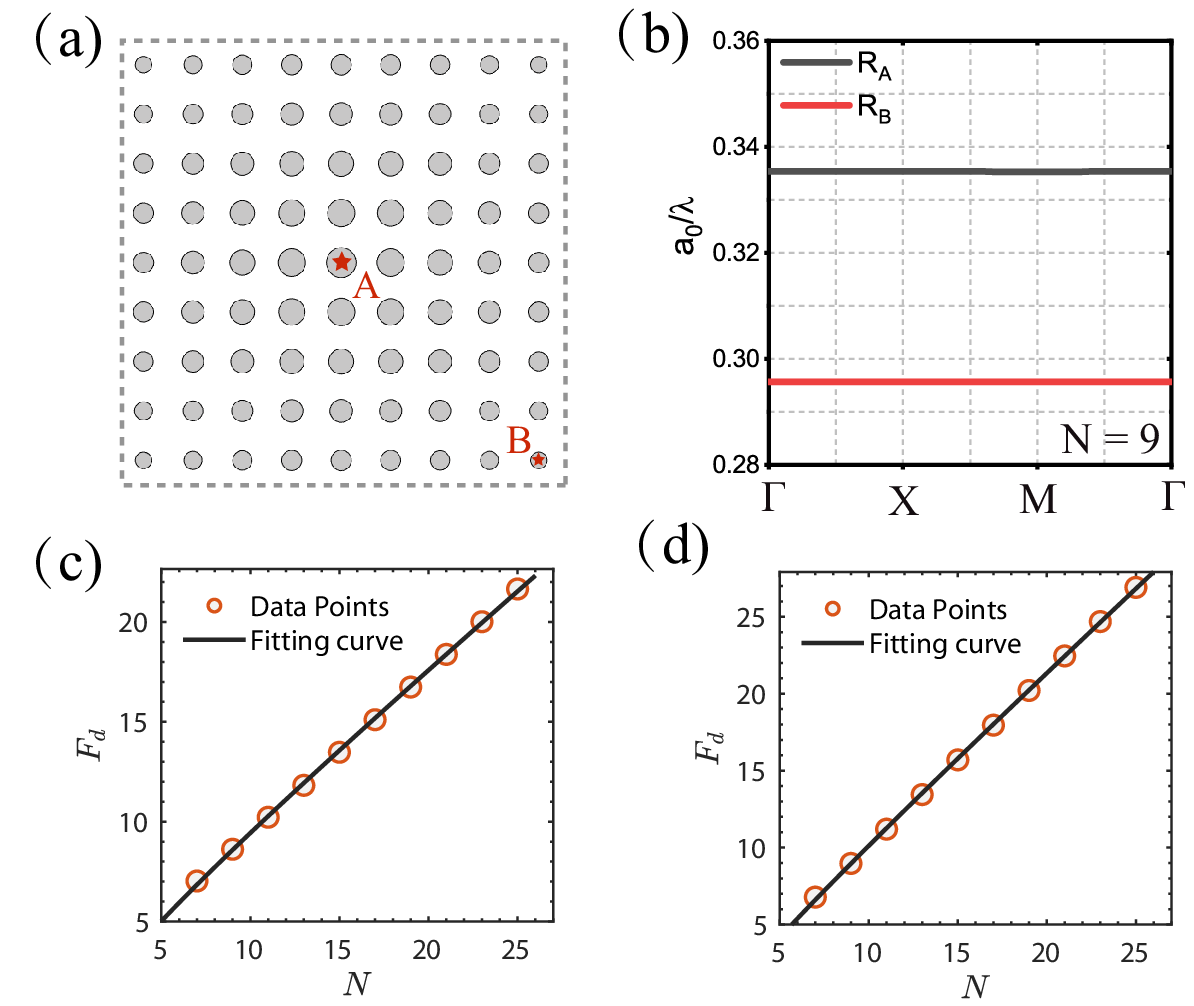}
	\caption{(a) A two dimensional moir\'e superlattice by employing spatially continuous perturbations to a perfect square lattice, wherein the rod diameter decays linearly.
	(b) Two flatbands $R_A$ and $R_B$ for the superlattice size $N$ = 9, where the field of $R_A$ and $R_B$ modes are localized around $A$ and $B$ sites, respectively. 	
	(c, d) The scaling of flatness $F_d$ on the supercell size $N$ for the flatbands $R_{A}$ (c) and $R_{B}$ (d), respectively. The data points are obtained via full-wave simulation, while fitting the data points with Eq.~\ref{eq:F_dN} gives rise to the fitting curves.	\label{Figure_4}}
\end{figure}

To further verify our theory, we design a two-dimensional moir\'e lattice as shown in Fig.~\ref{Figure_4}(a).
This moir\'e lattice is built on an ideal periodic lattice with square unit cell.
For this perfect lattice, the cell length is $a$, and each cell has a rod with diameter $d_0 (=0.6 a)$.
The continuous structure perturbations are implemented by introducing a position dependent decay of the rod diameter, i.e. $d(x,y) = d_0 - \gamma \sqrt{x^2 + y^2}$ where $\gamma = 0.3\sqrt{2}a/a_m$ with $a_m = Na$.
This moire lattice exhibit typical staggered band structure between the A and B sites, leading to wave localization either at the A site or B site.
Their corresponding flatbands are shown in Fig.\ref{Figure_4}(b) for supercell size $N = 9$.  
Again, we have compute moir\'e flatbands with different $N$, from which we obtain the dependence of the flatness $F_d$ on $N$ by using Eq.~\ref{eq:F_d_definition}.
The results for different flatbands are shown in Figs.~\ref{Figure_4}(c,d), respectively. It is evident that even with varied dimensions, the dependence of $F_d$ on $N$ consistently follows the scaling behavior predicted by Eq.~\ref{eq:F_dN}.
Similarly, the slopes for the two flatbands are quite different, which are 0.75 and 1.05, respectively. 
Based on our theory, the extracted Thouless localization length is $1.33a$ for the A-site localized mode, and is $0.95a$ for the B-site localized mode. 

In conclusion, we have demonstrated that the flatness of moir\'e bands is significantly modulated by the supercell size, and we have established a universal scaling law governing this relationship. By leveraging the statistical equivalence between moir\'e perturbations and disorder perturbations, we developed a rigorous scaling theory and derived its analytical expression. The accuracy of this scaling law is further validated through comprehensive full-wave simulations. Our findings enable the design of moir\'e systems via scales, shifting the paradigm from searching for magic conditions to leveraging the supercell size as a robust and predictable design knob. Consequently, these results offer a new framework for designing moir\'e resonant devices, with broad implications for quantum light sources~\cite{wang2025moire,yan2025cavity}, nonlinear optics~\cite{hong2022flatband,wang2024experimental,zhang2022ultralow}, and on-chip waveguiding\cite{wang2020localization,arkhipova2023observation}.

\vspace{12pt}
\begin{acknowledgments}
The authors wish to thank the support from National Key Research and Development Program of China (2022YFA1404604), National Natural Science Foundation of China (12574320), Fundamental Research Funds for the Central Universities(63261047), and Program for Innovative Research Team in Anqing Normal University.
\end{acknowledgments}

\bibliography{reference}

\end{document}